\documentclass[aps,twocolumn, pre,longbibliography, superscriptaddress]{revtex4-2}
\usepackage{graphicx}
\usepackage{epstopdf, epsfig}
\usepackage{amsbsy}
\usepackage{dutchcal} 

\usepackage{hyperref}
        \hypersetup{colorlinks=true, urlcolor=blue}

\usepackage{siunitx}

\usepackage[export]{adjustbox}

\usepackage[normalem]{ulem}
\usepackage{cancel}
\usepackage[utf8]{inputenc}
\AtBeginDocument{\renewcommand{\omega}{V}}

\usepackage{xcolor}
\usepackage{float, subcaption, caption}
\usepackage[english]{babel}
\usepackage{lipsum}
\usepackage{amssymb,amsfonts,amsmath}

\usepackage{mathtools}
\makeatletter
\newcommand\footnoteref[1]{\protected@xdef\@thefnmark{\ref{#1}}\@footnotemark}
\makeatother

\newcommand{\tnab}{\boldsymbol{\nabla}}
\newcommand{\tv}{\mathbf{v}}

\newcommand{\calH}{\mathcal{H}}

\newcommand{\Rd }{\mathcal{R} }

\newcommand{\uz }{\textbf{e}_z }

\newcommand{\tkm}{\mathbf{k}_m}
\newcommand{\tkn}{\mathbf{k}_n}
\newcommand{\tk}{\mathbf{k}}
\newcommand{\tx}{\mathbf{x}}

\newcommand{\taun}{\boldsymbol{\tau}_n}
\newcommand{\taum}{\boldsymbol{\tau}_m}

\newcommand{\ddX}{\dfrac{\partial}{\partial X}}

\newcommand{\ddY}{\dfrac{\partial}{\partial Y}}
\newcommand{\ddYsq}{\dfrac{\partial^2}{\partial Y^2}}

\newcommand{\m }{\text{m} }

\newcommand{\ti}{\mathcal{i}}

\newcommand{\ve }{\varepsilon}

\newcommand{\ux }{\textbf{e}_x }
\newcommand{\uy }{\textbf{e}_y }

\usepackage{bigints}

\makeatletter
\newcommand{\customlabel}[2]{%
\protected@write \@auxout {}{\string \newlabel {#1}{{#2}{}}}}
\makeatother

\usepackage{xcolor}

\begin{document}
\title{Fluid flow and spatiotemporal chaos in chemically active emulsions}

\author{Charu Datt}
\email{charudatt@mech.keio.ac.jp}
\affiliation{Max Planck Institute for the Physics of Complex Systems, N{\"o}thnitzer Stra{\ss}e 38, 01187 Dresden, Germany}
\affiliation{Department of Mechanical Engineering, Keio University,
Yokohama 223-8522, Japan}

\author{Jonathan Bauermann}
\affiliation{Max Planck Institute for the Physics of Complex Systems, N{\"o}thnitzer Stra{\ss}e 38, 01187 Dresden, Germany}
\affiliation{Department of Physics, Harvard University, Cambridge, MA 02138, USA}
\author{Nazmi Burak Budanur} 
\affiliation{Max Planck Institute for the Physics of Complex Systems, N{\"o}thnitzer Stra{\ss}e 38, 01187 Dresden, Germany}

\author{Frank J\"{u}licher} 
\affiliation{Max Planck Institute for the Physics of Complex Systems, N{\"o}thnitzer Stra{\ss}e 38, 01187 Dresden, Germany}
\affiliation{Center for Systems Biology Dresden, Pfotenhauerstra{\ss}e 108, 01307 Dresden, Germany}
\affiliation{Cluster of Excellence Physics of Life, TU Dresden, 01062 Dresden, Germany}

\begin{abstract}

We study phase-separating fluid mixtures as they demix in the presence of chemical reactions that maintain them away from thermodynamic equilibrium. We show that in such chemically active emulsions the interplay of chemical reactions, phase separation, and hydrodynamics effects complex self-organisation and pattern formation that can give rise to spatiotemporal chaos. This chaotic dynamics, unlike in classical turbulence, is not due to fluid inertia \textemdash we analyse the system in the Stokes flow regime \textemdash and it is different from the \emph{turbulence} of active nematics at low Reynolds number, for our fluid mixtures lack any orientational order. To explore the generic features of nonlinear
dynamics in our system, we derive amplitude equations which we find to be identical to those obtained for Rayleigh-Benard convection with mean flow and stress-free conditions at the top and bottom plates. Chemically active emulsions possessing no internal order, we thus establish, can exhibit chaoticity that is driven by interfacial stresses in the fluid mixture.

\end{abstract}

\maketitle

Active emulsions are phase-separating mixtures that are maintained away from thermodynamic equilibrium by energy input at small scales \cite{cates2018theories, Weber_2019}. Unlike passive emulsions \citep{siggia1979, bray2002theory} which typically reach a thermodynamically equilibrated macrophase-separated state, active emulsions can display intriquing non-equlibrium dynamics such as suppression and even reversal of the ripening process \citep{zwicker2015, Tjhung2018}. Active emulsions also provide a general framework to study the principles of self-organisation of multicomponent fluids inside living cells. Cells need to organise complex biochemical reactions within their confines \citep{hyman2014, Berry_2018, Weber_2019} and do so by creating distinct chemical environments in the form of compartments. Some of these compartments, such as the mitochondria --``the powerhouse of the cell'' \citep{siekevitz1957powerhouse}, have membranes, while others, like centrosomes which control shape, division and motility of the cells \citep{glover1993centrosome}, do not. Why then the compartments without membranes just not mix with their surroundings? This is because many membraneless organelles are, in fact, liquid droplets formed via phase separation inside cells \citep{brangwynne2009germline, hyman2014}. Their liquid-like properties such as viscosity and surface tension have been measured experimentally in several cases \citep{Jawerth2020,Wang2021}. In contrast to droplets in passive emulsions though, droplets inside cells are present in non-equilibrium environment where chemical reactions can actively regulate their size and composition.

The unravelling of liquid-liquid phase separation in biology requires understanding the coupled physics of active chemical reactions and phase separation. The interplay between them has indeed been a focus of current research \citep{zwicker_reaction}. In this context, however, hydrodynamic flows have often been neglected, or in other words, the viscosity of the phase-separating fluid mixture assumed infinite. Which begs the question: How does the viscosity of a chemically active phase-separating mixture affect its dynamics?   After all, we know that in passive phase separation, viscosity alters the kinetics of the separation process and the morphology of its transient structures \citep{siggia1979, datt_2015}. The motivation behind the present article is to explore the importance of fluid mechanics in phase-separating mixtures with active chemical reactions.

We study an incompressible binary fluid mixture with components A and B which undergo reversible chemical reactions $\mathrm{A} \rightleftharpoons \mathrm{B}$ with constant forward and backward reaction rates $\Gamma_f$ and $\Gamma_b$, respectively. Having constant reaction rates in a phase-separating system is inconsistent with the thermodynamic condition of detailed balance \cite{Glotzer_original, Lefever_comment, Glotzer_reply} and implies an active system (\emph{vid.\ }\citep{vrugt2025whatexactlyisactivematter}). An important remark is already in place here. When fluid flows are neglected, as they are in many previous studies on active emulsions, the coarse-grained model of the binary active emulsion that we have just described is equivalent to the model of spinodal decomposition of diblock copolymers, which is an equilibrium problem \citep{Leibler1980, glotzer1994_block, Sagui1994}.  This mapping onto an equilibrium system breaks down when fluid flows are accounted for, thereby highlighting the flows' importance in effectuating the non-equilibrium dynamics.

In this Letter,  we highlight the most intriguing aspects of our minimal system, a binary mixture with the simplest of chemical reactions. We show that active chemical reactions drive persistent flows. When these flows are weak, e.g. for low reactions rates and high viscosities, the system reaches a steady state that is either stationary or travelling with a constant drift velocity \cite{goldstein1991}. When the flows are strong, the system exhibits chaotic dynamics. This chaotic dynamics is novel in that it is neither due to fluid inertia nor like the turbulence of active nematics: we analyse the system in the Stokes flow regime and contra nematics, our system lacks any orientational order and is a \emph{scalar} mixture.

\emph{Minimal model of an active emulsion with flows} \textemdash The dynamics of passive phase separation of an incompressible binary fluid is governed by model H of Hohenberg and Halperin \citep{hohenberg1977, cates2018theories}.  The model comprises of the Navier-Stokes equations for fluid momentum and mass conservation, and an equation for conserved order parameter $\phi$. We choose $\phi$ as the local molecular composition of the mixture: $\phi = \langle n_A - n_B\rangle/ \langle n_A + n_B\rangle$, where $n_{A, B}$ represents the concentration of molecules of A, B \cite{cates2018theories}. We describe the phases using the Ginzburg-Landau free-energy \cite{cates2018theories, bray2002theory}
 ${F}[\phi] \equiv \bigintssss (V \left(\phi \right) + {\kappa} \left( \nabla \phi\right)^2/2)\, d\text{V}$ with
$ V \left(\phi \right) \equiv -{a} \phi^2/2 + {b} \phi^4/4  $, which has a double-well structure with minima at $\pm \phi_b$ where $\phi_b \equiv \sqrt{a/b}$, and $a$, $b$ and $\kappa$ are positive constants.
 The equations that we study are
\begin{align}
&\dfrac{\partial \phi}{\partial t} + {\mathbf{v}\cdot \boldsymbol{\nabla} \phi} = {M}\;\nabla^2 \mu  -  {R} \left(\phi - \phi_s \right),
\label{phi_eq}\\
-&\nabla p +\eta \nabla^2 \mathbf{v}  =   {\phi \boldsymbol{\nabla} \mu}\quad \text{ with } \quad \boldsymbol{\nabla} \cdot \mathbf{v} = 0,
\label{momentum_eq}
\end{align}
where, expecting small inertial forces compared to viscous forces at the scale of biological cells, we have considered only the Stokes equations. The fluid components A and B here have, for simplicity, the same mass density $\rho$ and viscosity $\eta$ but an interfacial tension $\gamma \equiv \sqrt{8\kappa a^3/9b^2}$ between them.  The field $\phi$ is convected by the flow velocity ${\bf v}$, and $p$ is the Lagrange multiplier imposing
incompressibility of the flow field, $\boldsymbol{\nabla} \cdot \mathbf{v} = 0 $. The chemical potential is defined as $\mu \equiv {\delta {F}}/{\delta \phi}$ and the mobility $M$ is chosen to be constant. The force density -$\phi\nabla\mu$ equals the divergence of the equilibrium
stress stemming from interfacial tension between the two fluids, in accordance with the Gibbs-Duhem relation. The chemical reactions make the system active and drive it towards a spatially-averaged value of $\langle \phi \rangle = \phi_s$, where $\phi_s \equiv \left(\Gamma_b - \Gamma_f\right)/ {R} $ and ${R} \equiv \Gamma_f + \Gamma_b$ \cite{Huo2003}.

It is instructive to work with dimensionless quantities: we therefore scale lengths with $\ell \equiv  w/\sqrt{2}$ where $w \equiv \sqrt{2 \kappa/ a}$ characterizes the interface width, time with $\tau \equiv w^2/D$ where $D \equiv 2 a M$ is the effective diffusion constant at late times \cite{bray2002theory}, and $\phi$ by $\phi_b$ to obtain
\begin{align}
\dfrac{\partial \phi}{\partial t} + {\mathbf{v}\cdot \boldsymbol{\nabla} \phi} &=\nabla^2 \left(-\phi + \phi^3 - \nabla^2 \phi\right)  -  \mathcal{R} \left(\phi - {m}\right),
\label{phi_eq_nonD}\\
-\nabla p + \nabla^2 \mathbf{v}  &=  \mathcal{H} {\phi \boldsymbol{\nabla} \left(-\phi + \phi^3 - \nabla^2 \phi\right)}
\label{momentum_eq_nonD}
\end{align}
along with $\boldsymbol{\nabla} \cdot \mathbf{v} = 0$, where the variables now and henceforth represent scaled and dimensionless quantities.
The dimensionless parameters $\mathcal{R} \equiv R \tau$, $\mathcal{H} \equiv 3 \gamma \tau/2 \eta w$, and $m \equiv \phi_s/\phi_b$ represent the strength of activity, of hydrodynamic flows, and of asymmetry in the composition, respectively, and control the dynamics of the system.
We solve the equations numerically using the Fourier pseudospectral method in two spatial dimensions with a square box of side $256\sqrt{2}$ and  periodic boundary conditions. The number of grid points in each spatial direction is initially $512$, in other words two points per interface width, and $1024$ at late times, including when quantifying chaos. The time integration is performed using the Runga-Kutta (2, 3, 3) implicit-explicit scheme \cite{ascher1997}. As we have a cubic non-linearity in the equations, we use 1/2-dealiasing (truncation) scheme to eliminate the aliasing error \citep{pandit2024, dealias_paper}. We initialise the simulations with a fully mixed state of $\phi = m$ plus a small random noise to destabilise it.

\emph{Results and discussion} \textemdash
Even as phase separation tends to demix the components, the chemical reactions mix them.  As we stated above, the active emulsion model of \eqref{phi_eq} and \eqref{momentum_eq} but without hydrodynamic flows, i.e.\ $\mathbf{v} = \mathbf{0}$, is equivalent to an equilibrium model for diblock copolymers with competing local and nonlocal interactions \citep{glotzer1994_block}. This system will eventually reach an equilibrium state. In two spatial dimensions, the global free-energy minima of the hydrodynamics-absent system  can be (i) a homogeneous state with uniform concentration $\phi=m$, (ii)
periodic stripe patterns and (iii) periodic  patterns with hexagonal symmetry \citep{rustum2011}. Typically, small values of $\vert m \vert$ lead to stripes, large values close to the value $\vert m \vert = 1$ lead to uniform states. Hexagonal patterns
are stable for intermediate values of $m$ \cite{rustum2011}. The charateristic size of the patterns is inversely proportional to the reaction rate \citep{glotzer1995, Glotzer_original, Liu1989}. When hydrodynamic flows are present, the active emulsion model can no longer be mapped onto the equlibrium problem due to the different definition of chemical potential in the active emulsion and diblock-copolymer models. Since the gradient of chemical potential enters the Stokes equation, the fluid flows in the two systems will be different.

In order to explore the effect of fluid flows, we consider a wide range of viscosity values such that $\mathcal{H}$ ranges from 0.05 to 500, three different values for the strength of reactions $\Rd$, namely, 0.008, 0.04, 0.08, and two values of $m$, 0 and -0.2 corresponding to {50:50} and {40:60} component ratio, respectively. These  values of $\Rd$ correspond to molecular diffusion across the interface being faster than the interconversion rate of the components. Physical systems such as oil-water mixtures will have $\calH$ around $10$ with $w \approx 1$ \si{\nano \meter}, $\gamma \approx 10^{-2}$ \si{\newton\per\meter}, $\eta \approx 10^{-3}$ \si{\newton\second\meter^{-2}}, and $D \approx 10^{-9}$ \si{\meter^2 \per\second}. For soft colloidal liquids or biocondensates inside biological cells, it may vary from $10^{-4}$ to $0.1$ \cite{Seyboldt_2018}. It is important to remember, however, that intracellular mixtures undergoing phase separation can hardly be called binary and enzymatic chemical reactions are rarely as simple; our motivation in studying the minimal model is only to uncover the basic principles that may underlie real systems.
\begin{figure*}
\centering
    \includegraphics[width = 01\textwidth]{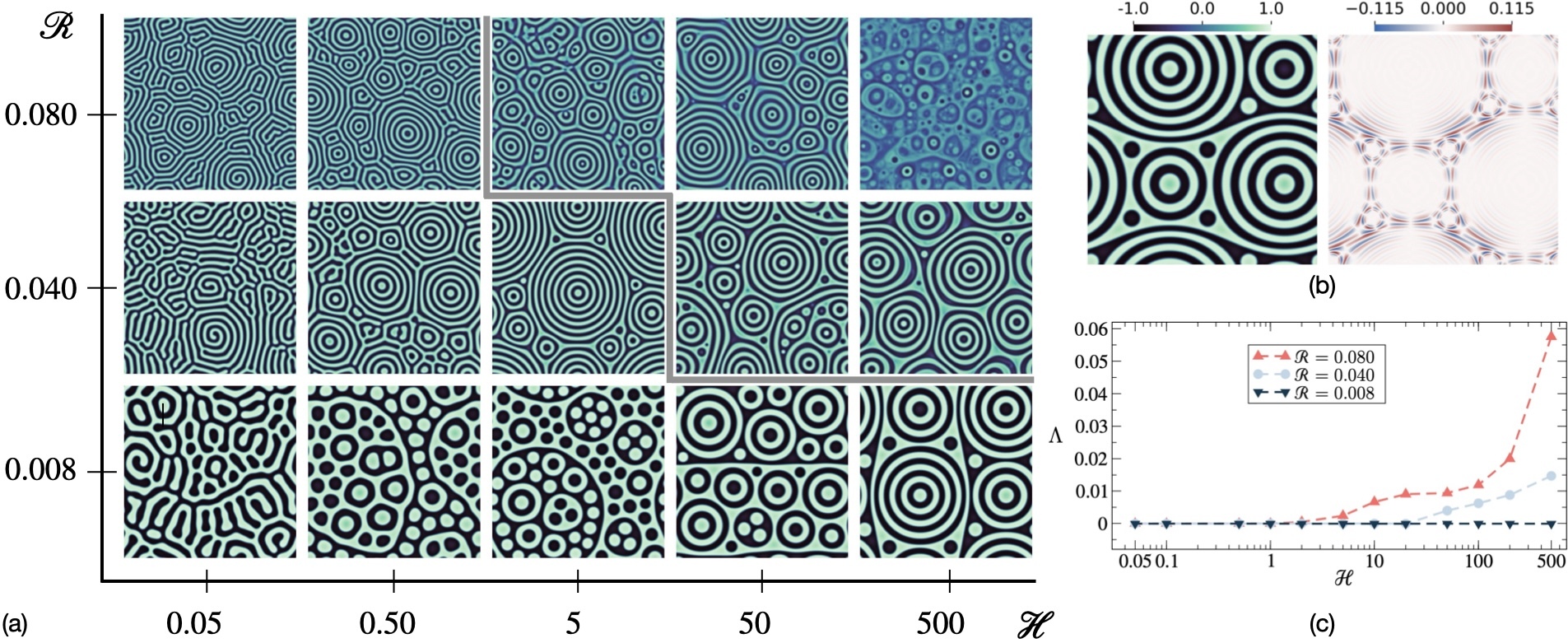}
    \caption{For symmeteric mixtures. a) Patterns in $\phi$ after long times of $\mathcal{O}\left(10^5\right)$ for some values of $\calH$ and $\Rd$. The grey line separates the steady patterns from chaotic ones. Colorbar same as in b).  b) A steady $\phi$ pattern (left) and the underlying vorticity (right) for $\Rd = 0.008$ and $\calH = 200$. c) Plot showing values of the maximal Lyapunov exponent for different $\calH$ and $\Rd$. Contrary to the dynamics of Swift-Hohenberg equation with mean flow \citep{karimi2011, Chaim2003} \textemdash a system related to Rayleigh-Benard convection, we find the spiral defects to be stable whereas it is the target patterns that populate the chaotic state.}
    \label{fig:5050summary}
\end{figure*}

The presence of fluid flows brings in fundamentally new dynamics. For low values of $\calH$ and $\Rd$, the system reaches a steady-state with patterns whose morphology is dependent on these values. For symmeteric mixtures ($m = 0$), in addition to the lamellae and stripes of the no-hydrodynamic case we now also find spiral and target patterns (see figure \ref{fig:5050summary}a), as was previously reported in \cite{Huo2003, furtado2006} for chemically reactive binary mixtures. Fluid flows underpin these steady patterns, as can be seen in figure \ref{fig:5050summary}b. If the hydrodynamics were to be turned off, the patterns would become unstable and decay towards the stripe pattern which is the global energy minimizer in the absence of hydrodynamics in symmetric mixtures. Importantly, however, not every trajectory of the system leads to a steady state. At high $\calH$ and $\Rd$, the patterns show chaotic dynamics, which we ascertain by calculating the maximal Lyapunov exponent $\Lambda$. Positive values of $\Lambda$ indicate unpredictability and in figure \ref{fig:5050summary}c we find these for high values of $\calH$ and $\Rd$. In brief, we use the procedure described in \cite{benettin1976} to calculate the exponent. We begin the procedure after pattern statistics have fully developed, and denote with $\phi \left(0\right)$ the state of the concentration field at the procedure's start; $\phi\left(t\right)$ then represents the state after time $t$, $\phi \left(t\right) = \mathcal{Q}^t \left[\phi \left(0\right) \right]$, where $\mathcal{Q}^t$ is the time-$t$ map due to the dynamics of the system. Next, we add a random perturbation $\delta \phi\left(0\right)$ to the initial state $\phi \left(0\right)$ according to $\vert \vert \delta \phi\left(0\right)\vert \vert = \sigma \vert \vert\phi \left(0\right)\vert \vert$, where $\sigma$ is a small real positive value and $\vert\vert \cdot \vert \vert$ represents the $L_2$ norm. We then march $\phi \left(t\right)$ and $\phi \left(t\right) + \delta \phi\left(t\right) $ in parallel in time. After every time span of $\tau_{\Lambda}$,  we rescale the perturbation such that $\vert \vert \delta \phi\left(n \tau_{\Lambda} \right)\vert \vert = \sigma \vert \vert\phi \left(n \tau_{\Lambda}\right)\vert \vert$, where $n = 1, 2,3 \ldots$. The maximal Lyapunov exponent is then estimated as $\Lambda \approx \frac{1}{N_{\Lambda} \tau_{\Lambda}}\Sigma_{n=1}^{N_{\Lambda}} \ln{\dfrac{\vert \vert \mathcal{Q}^{\tau_{\Lambda}}\left[\phi \left(n \tau_{\Lambda}\right) + \delta \phi \left(n \tau_{\Lambda}\right)  \right] - \mathcal{Q}^{\tau_{\Lambda}}\left[ \phi \left(n \tau_{\Lambda}\right) \right]  \vert \vert}{\vert \vert \delta \phi\left(n \tau_{\Lambda} \right)\vert \vert}}$. We have used $\tau_{\Lambda} = 0.25$ and  $\sigma= 10^{-4}$ and verified that these parameters had no significant bearing on the results by increasing and decreasing $\sigma$ by an order of magnitude, and halving and doubling the value of $\tau_{\Lambda}$. $N_{\Lambda}$ was chosen as $2.4 \times 10^5$ and in figure \ref{fig:5050summary}c, we plot the average values and errorbars of $\Lambda$ for the last $1.2 \times 10^5$ rescalings. We did not see any long time trends in the exponent values, and the errorbars in the figure are smaller than the symbol size, indicating convergence of values.

\begin{figure}
\centering
    \includegraphics[width = 0.49\textwidth]{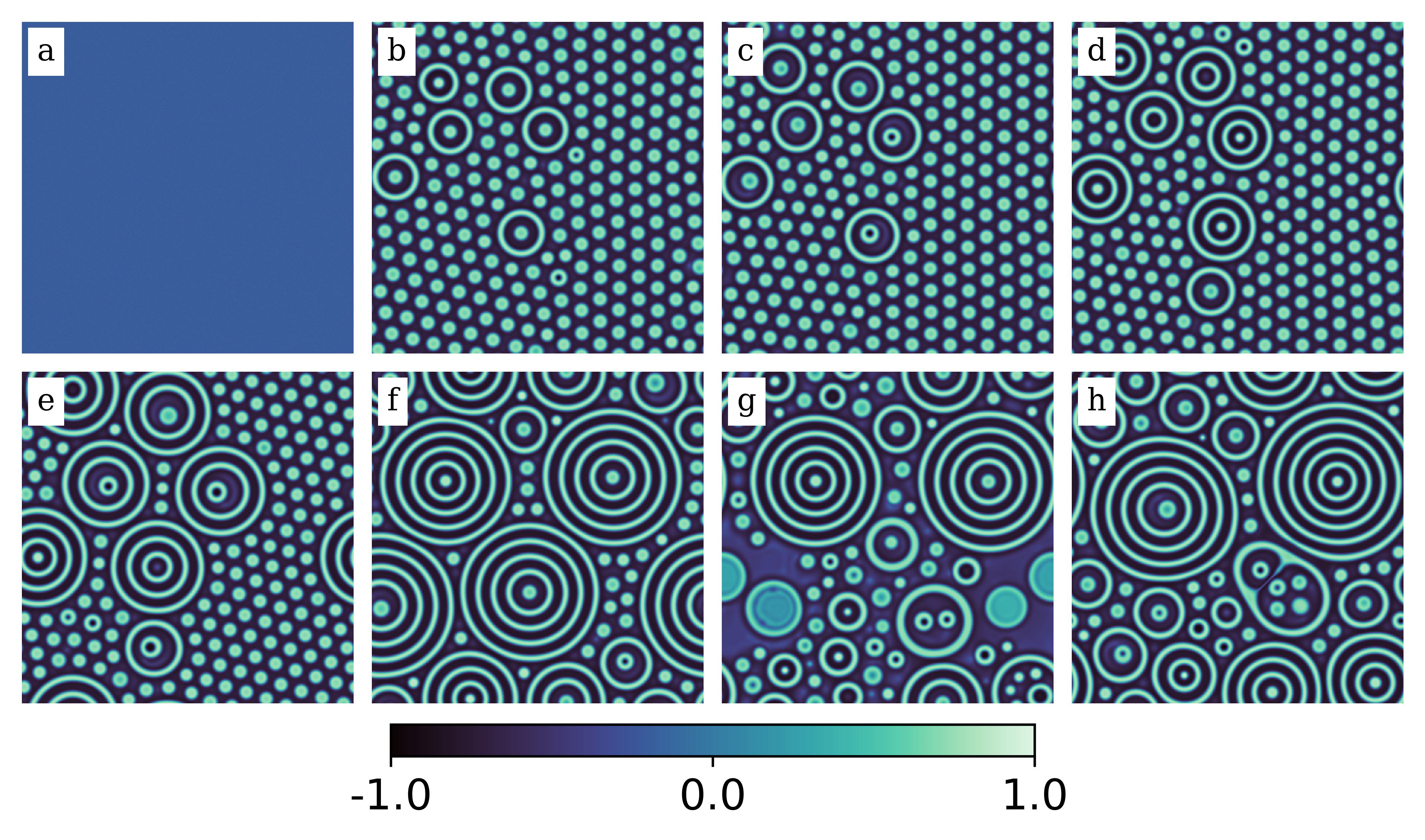}
    \caption{Enroute to spatio-temporal chaos in a 40:60 mixture for $\calH = 500$ and $\Rd = 0.04$. Snapshots of the $\phi$ field at a) $t = 0$ b) $t = 1250$ c) $t = 2500$ d) $t = 5000$ e) $t = 10000$  f) $t = 19500$ g) $t = 19750$ h) $t = 25000$. The Lyapunov exponent $\Lambda = 0.00529$ at late times of $\mathcal{O}\left(10^6\right)$.}
    \label{fig:40-60chaos}
\end{figure}

While the ensuing fluid flows are stronger in symmetric mixtures owing to the continuous nature of the phase domains \citep{siggia1979}, we also observe chaotic dynamics for asymmetric mixtures. In figure \ref{fig:40-60chaos}, we highlight the dynamics of a 40:60 mixture. Here, we initially see double phase separation \citep{tanaka1995} and shell formation in a few drops. These shells then grow adding layers onto themselves from the material of the surrounding drops, collide, collapse and reform, and so the process continues. Conversely, in the absence of hydrodynamics, the pattern would be one of arrested drops, ideally on a hexagonal lattice \citep{rustum2011}.

With the aim of understanding the generic features of the dynamics of our model, we
derive amplitude equations which describe the effects of non-linearities at
large scales and long times on some simple steady-state patterns. In particular, we focus on stripes and hexagonal spots that appear in the absence of hydrodynamics as noted before, and show how hydrodynamics can affect the evolution of these  near the threshold of pattern-forming instability.

We study the system near the threshold of instability which occurs when the reaction-diffusion lengthscale (dimensional) $\xi \equiv \left(D/R\right)^{1/2}$ becomes  similar to the length scale (dimensional) $w$ characterizing the interface width
of the phase-separating domains.
We proceed by defining $u \equiv \phi - \m$ in \eqref{phi_eq_nonD} and \eqref{momentum_eq_nonD}, and use the stream function $\psi$ to write
$\tv \left(x, y\right) = \tnab \psi \times \uz$, exploiting the two-dimensional spatial nature of the problem.

We first discuss the linear stability of the uniform state, $u = 0$. For $u = \hat{u} \exp {\left(\ti \tk \cdot \tx + \lambda t\right)}$, the dispersion relation is
\begin{equation}
 \lambda(k) = (1 - 3\m^2 ) k^2 - k^4 - \Rd \quad .
\end{equation}
The wave vector $k_c$ for which $d\lambda/dk=0$, and therefore $k_c^2=(1-3m^2)/2$, gives the fastest growing mode provided that $\vert m \vert < 1/\sqrt{3}$.  The instabiliy of this mode occurs at $\Rd=\Rd_c$ where $\Rd_c=k_c^4$, so that  $\lambda = - \left(k^2 - k_c^2 \right)^2 +\Rd_c-\Rd$. We study the dynamics near this bifurcation point, i.e.
when $\Rd_c-\Rd= r \varepsilon^2$, where $\varepsilon$ is $\mathcal{o}\left( 1\right)$ and $r$ is the control parameter \cite{hoyle2006pattern,shiwa1997amplitude}. 

We then proceed systematically to derive the amplitude equations \cite{gunaratne1994pattern, hoyle2006pattern}. Looking first at the slow dynamics of stripe patterns
 aligned perpendicular to, say the x-axis,  
 we write
 \begin{equation}
 \begin{aligned}
 u(x,y,t)  &=  \epsilon A \left(\ve x, \ve^{1/2} y, \ve^2 t \right) e^{\ti k_c x}  \text{  +  c.c.} + O(\ve^2) \\
 \psi(x,y,t)  &= \epsilon^{3/2} B \left(\ve x, \ve^{1/2} y, \ve^2 t \right) +O(\ve^{5/2})
 \end{aligned}
 \end{equation}
 where $A(X,Y,T)$ is complex and $B(X,Y,T)$ is real and $ X = \ve x, \: \: Y = \sqrt{\ve} y, \: \: T = \ve^2 t$ are the slow scales.
The dynamics of the amplitude $A$  at the lowest order in $\ve$ is then given by
\begin{equation}
{\small
\begin{aligned}
 &\dfrac{\partial A}{\partial T} = r A - 3 A \vert A\vert^2 k_c^2 + 4 k_c^2 \left(\dfrac{\partial}{\partial X}
 - \dfrac{i}{2k_c}\dfrac{\partial^2}{\partial Y^2}\right)^2 A - ik_c \dfrac{\partial B}{\partial Y} A\\
 &\frac{\partial^4B}{\partial Y^4} = {2 k_c^2} \calH \left[ \ddY \left(\bar{A} \left( \ddX - \dfrac{\ti}{2k_c} \ddYsq\right) A \right) \right] \text{   +   c.c.}
 \end{aligned}\
 }
 \label{eq:stripeAmpli}
 \end{equation}
where $\bar{A}$ is complex conjugate of $A$. These equations \eqref{eq:stripeAmpli} are identical in form to those obtained for the stability of rolls at low Prandtl number Rayleigh-Benard convection with stress-free boundary conditions \cite{hoyle2006pattern,siggia1981}.

Similarly, for hexagon patterns at leading order we write $u(\tx,t) = \sum_{n=1}^{3} \ve A_n \left(\ve \tx, \ve^2 t\right) e^{i {\bf k}_n \cdot \tx} \text{  +  c.c. }$ with $\psi(\tx,t)= \ve B(\ve \tx, \ve^2 t)$
where $\tk_1 = k_c \ux$, $\tk_2 = k_c \left( -\ux +\sqrt{3} \uy \right)/2$ and $\tk_3 = k_c \left(-\ux -\sqrt{3}  \uy \right)/2$. For a systematic derivation now, however, we must substitute $3m =\ve s$ and $\calH = \ve h$ where $s$ and $h$ are constants  \citep{Matthews1998, young_mean_flow}.  The resulting amplitude equations are
{\small \begin{equation}
\begin{aligned}
 \frac{\partial A_n}{\partial T} &= r A_n + 4 \left(\tkn \cdot \nabla \right)^2 A_n
 - 2 s k_c^2 \left[\bar A_{n-1} \bar A_{n+1} \right]
 \\
        &  -3k_c^2 \left[\vert A_n \vert^2  + 2\vert A_{n+1}\vert^2  + 2\vert A_{n-1}\vert^2 
          \right] A_{n}  -\ti A_n \taun \cdot \nabla B\\
\nabla^4 B&= \sum_{m=1}^3{2} h \left[\left(\taum \cdot \nabla\right) \left(\tkm \cdot \nabla\right)\vert A_m\vert^2 \right],
\end{aligned}
\label{eq:hexAmpli}
\end{equation}}
where $n-1$ and $n+1$ are understood modulo 3. $\nabla$  acts on $(X, Y)$ and $\taum$ are obtained from ${\bf k}_m$
by an anticlockwise rotation of $\pi/2$. Again, equations \ref{eq:hexAmpli} are of the same form as those obtained for the evolution of hexagon patterns in non-Bousinessq Rayleigh-Benard convection with mean flow \cite{young_mean_flow}.

These amplitude equations, unlike the Newell-Whitehead-Segel equation that governs the dynamics in the absence of hydrodynamic flows \cite{Newell_Whitehead_1969, Segel_1969, shiwa1997amplitude}, do not have a Lyapunov functional and therefore  can admit oscillatory solutions \cite{hoyle2006pattern}. The identical nature of Ginzburg-Landau type equations in Rayleigh-Benard and chemically active emulsions suggests that the latter also offer rich dynamical behaviour, and this behaviour can be further investigated through insights developed for the well-studied former problem.

\emph{Conclusion and outlook} \textemdash We studied binary fluid phase separation in the presence of active chemical reactions, and found that the reactions drive persistent fluid flows.  When these flows are weak, the system reaches a steady state; when strong, spatiotemporal chaos emerges. By deriving the amplitude equations, we showed that the weakly non-linear dynamics of our system is related to Rayleigh-Benard convection. However, in the numerical simulations of the governing equations, we did not observe the typical spiral defect chaos of the convection problem \cite{chiam2003, karimi2011, Bodenschatz2000}, instead the chaotic states were dominated by target patterns. We note that similar patterns and an unsteady dynamics were reported for phase-field crystal model (a conserved version of the Swift-Hohenberg equation) with mean flow \cite{shiwa2005a,shiwa_conference}, where unlike in our system a neutrally stable Goldstone mode is present (\emph{vid.} \citep{Ohnogi2011}). Binary fluid mixtures de-mixing under gravity and a continuously ramped temperature that creates a buoyancy effect have also shown oscillatory dynamics \citep{cates2003}. The similarities and differences of these systems inspire our future work.

It is important that we contrast our system with others that are similar. Previously, in \cite{golovin2001, golovin2004} \textit{convective} Cahn-Hilliard models of the form $\phi_t + {\mathbf{v}\cdot \boldsymbol{\nabla} \phi} = \nabla^2 \mu$ were considered, where the velocity $\mathbf{v} $ was assumed to be $G \nabla \phi$ with the constant $G$ representing the intensity of a driving force.  The authors of these studies found that while the system showed coarsening dynamics for weak driving $G\rightarrow 0$,  as $G \rightarrow \infty$ chaotic structures typical of Kuramoto-Sivashinsky equation emerged. In the present work, we do not assume a form of velocity field but instead solve the fully coupled Stokes-Cahn-Hilliard equations driven via active chemical reactions. Chemical reactions have also been known to drive hydrodynamic flows in various non-phase-separating systems, for example, in the canonical Belousov—Zhabotinskii reaction \cite{Miike1988, Miike1989}. In these systems, however, the chemical reactions contribute to fluid momentum through local variations of bulk quantities like fluid density and viscosity \cite{Diewald1995, DeWit_review, Huffman2023}. This is different to the case we considered here where reactions contribute to fluid forces via constant generation of interfacial stresses, through the term $\phi \nabla \mu$ in the Stokes equation. Additionally, our model is still different, both in spirit and form, to the active phase-separating model Active Model H that presents a continuum theory for self-propelled particles \cite{tiribocchi2015}. In that model, although activity manifests itself at interfaces, it is added phenomenologically through terms that break time reversal symmetry in chemical potential and fluid stresses and are different to those that emerge here due to chemical activity.

In the end, it is worth emphasizing that we have studied only a minimal model of active emulsions. We did not consider fluid inertia
along with any differences in the density and viscosity of the components, all of which can generate richer dynamics \citep{lowengrub1978,Onuki_1994,cates2003, Huo2004}. We ignored the Langevin noise of the original model H \citep{hohenberg1977, cates2018theories}, when noise in active emulsions may play an important role. Moreover, many real systems are multicomponent systems which behave fundamentally differently from binary mixtures \citep{bauermann2022chemical}. While biological emulsions are often rheologically complex \citep{Jawerth2020, tanaka2022viscoelastic}, showing properties such as viscoelasticity, which brings in a new time scale to the problem \citep{Tanaka2000}, as a first step, we restricted our analysis to only Newtonian fluids. We believe that this work will stimulate further exploration of the hydrodynamics of active emulsions.

\emph{Acknowledgements}\textemdash C.D. is grateful to Alexander Morozov of University of Edinburgh for discussions. All authors acknowledge support from the Max Planck Society.

\bibliography{phaseSeparation}

\begin{thebibliography}{65}%
\makeatletter
\providecommand \@ifxundefined [1]{%
 \@ifx{#1\undefined}
}%
\providecommand \@ifnum [1]{%
 \ifnum #1\expandafter \@firstoftwo
 \else \expandafter \@secondoftwo
 \fi
}%
\providecommand \@ifx [1]{%
 \ifx #1\expandafter \@firstoftwo
 \else \expandafter \@secondoftwo
 \fi
}%
\providecommand \natexlab [1]{#1}%
\providecommand \enquote  [1]{``#1''}%
\providecommand \bibnamefont  [1]{#1}%
\providecommand \bibfnamefont [1]{#1}%
\providecommand \citenamefont [1]{#1}%
\providecommand \href@noop [0]{\@secondoftwo}%
\providecommand \href [0]{\begingroup \@sanitize@url \@href}%
\providecommand \@href[1]{\@@startlink{#1}\@@href}%
\providecommand \@@href[1]{\endgroup#1\@@endlink}%
\providecommand \@sanitize@url [0]{\catcode `\\12\catcode `\$12\catcode
  `\&12\catcode `\#12\catcode `\^12\catcode `\_12\catcode `\%12\relax}%
\providecommand \@@startlink[1]{}%
\providecommand \@@endlink[0]{}%
\providecommand \url  [0]{\begingroup\@sanitize@url \@url }%
\providecommand \@url [1]{\endgroup\@href {#1}{\urlprefix }}%
\providecommand \urlprefix  [0]{URL }%
\providecommand \Eprint [0]{\href }%
\providecommand \doibase [0]{https://doi.org/}%
\providecommand \selectlanguage [0]{\@gobble}%
\providecommand \bibinfo  [0]{\@secondoftwo}%
\providecommand \bibfield  [0]{\@secondoftwo}%
\providecommand \translation [1]{[#1]}%
\providecommand \BibitemOpen [0]{}%
\providecommand \bibitemStop [0]{}%
\providecommand \bibitemNoStop [0]{.\EOS\space}%
\providecommand \EOS [0]{\spacefactor3000\relax}%
\providecommand \BibitemShut  [1]{\csname bibitem#1\endcsname}%
\let\auto@bib@innerbib\@empty
\bibitem [{\citenamefont {Cates}\ and\ \citenamefont
  {Tjhung}(2018)}]{cates2018theories}%
  \BibitemOpen
  \bibfield  {author} {\bibinfo {author} {\bibfnamefont {M.~E.}\ \bibnamefont
  {Cates}}\ and\ \bibinfo {author} {\bibfnamefont {E.}~\bibnamefont {Tjhung}},\
  }\bibfield  {title} {\bibinfo {title} {Theories of binary fluid mixtures:
  from phase-separation kinetics to active emulsions},\ }\href
  {https://doi.org/10.1017/jfm.2017.832} {\bibfield  {journal} {\bibinfo
  {journal} {J. Fluid Mech.}\ }\textbf {\bibinfo {volume} {836}},\ \bibinfo
  {pages} {P1} (\bibinfo {year} {2018})}\BibitemShut {NoStop}%
\bibitem [{\citenamefont {Weber}\ \emph {et~al.}(2019)\citenamefont {Weber},
  \citenamefont {Zwicker}, \citenamefont {Jülicher},\ and\ \citenamefont
  {Lee}}]{Weber_2019}%
  \BibitemOpen
  \bibfield  {author} {\bibinfo {author} {\bibfnamefont {C.~A.}\ \bibnamefont
  {Weber}}, \bibinfo {author} {\bibfnamefont {D.}~\bibnamefont {Zwicker}},
  \bibinfo {author} {\bibfnamefont {F.}~\bibnamefont {Jülicher}},\ and\
  \bibinfo {author} {\bibfnamefont {C.~F.}\ \bibnamefont {Lee}},\ }\bibfield
  {title} {\bibinfo {title} {Physics of active emulsions},\ }\href
  {https://doi.org/10.1088/1361-6633/ab052b} {\bibfield  {journal} {\bibinfo
  {journal} {Rep. Prog. Phys.}\ }\textbf {\bibinfo {volume} {82}},\ \bibinfo
  {pages} {064601} (\bibinfo {year} {2019})}\BibitemShut {NoStop}%
\bibitem [{\citenamefont {Siggia}(1979)}]{siggia1979}%
  \BibitemOpen
  \bibfield  {author} {\bibinfo {author} {\bibfnamefont {E.~D.}\ \bibnamefont
  {Siggia}},\ }\bibfield  {title} {\bibinfo {title} {Late stages of spinodal
  decomposition in binary mixtures},\ }\href
  {https://doi.org/10.1103/PhysRevA.20.595} {\bibfield  {journal} {\bibinfo
  {journal} {Phys. Rev. A}\ }\textbf {\bibinfo {volume} {20}},\ \bibinfo
  {pages} {595} (\bibinfo {year} {1979})}\BibitemShut {NoStop}%
\bibitem [{\citenamefont {Bray}(2002)}]{bray2002theory}%
  \BibitemOpen
  \bibfield  {author} {\bibinfo {author} {\bibfnamefont {A.~J.}\ \bibnamefont
  {Bray}},\ }\bibfield  {title} {\bibinfo {title} {Theory of phase-ordering
  kinetics},\ }\href {https://doi.org/10.1080/00018730110117433} {\bibfield
  {journal} {\bibinfo  {journal} {Adv. Phys.}\ }\textbf {\bibinfo {volume}
  {51}},\ \bibinfo {pages} {481} (\bibinfo {year} {2002})}\BibitemShut
  {NoStop}%
\bibitem [{\citenamefont {Zwicker}\ \emph {et~al.}(2015)\citenamefont
  {Zwicker}, \citenamefont {Hyman},\ and\ \citenamefont
  {J\"ulicher}}]{zwicker2015}%
  \BibitemOpen
  \bibfield  {author} {\bibinfo {author} {\bibfnamefont {D.}~\bibnamefont
  {Zwicker}}, \bibinfo {author} {\bibfnamefont {A.~A.}\ \bibnamefont {Hyman}},\
  and\ \bibinfo {author} {\bibfnamefont {F.}~\bibnamefont {J\"ulicher}},\
  }\bibfield  {title} {\bibinfo {title} {Suppression of {O}stwald ripening in
  active emulsions},\ }\href {https://doi.org/10.1103/PhysRevE.92.012317}
  {\bibfield  {journal} {\bibinfo  {journal} {Phys. Rev. E}\ }\textbf {\bibinfo
  {volume} {92}},\ \bibinfo {pages} {012317} (\bibinfo {year}
  {2015})}\BibitemShut {NoStop}%
\bibitem [{\citenamefont {Tjhung}\ \emph {et~al.}(2018)\citenamefont {Tjhung},
  \citenamefont {Nardini},\ and\ \citenamefont {Cates}}]{Tjhung2018}%
  \BibitemOpen
  \bibfield  {author} {\bibinfo {author} {\bibfnamefont {E.}~\bibnamefont
  {Tjhung}}, \bibinfo {author} {\bibfnamefont {C.}~\bibnamefont {Nardini}},\
  and\ \bibinfo {author} {\bibfnamefont {M.~E.}\ \bibnamefont {Cates}},\
  }\bibfield  {title} {\bibinfo {title} {Cluster phases and bubbly phase
  separation in active fluids: Reversal of the {O}stwald process},\ }\href
  {https://doi.org/10.1103/PhysRevX.8.031080} {\bibfield  {journal} {\bibinfo
  {journal} {Phys. Rev. X}\ }\textbf {\bibinfo {volume} {8}},\ \bibinfo {pages}
  {031080} (\bibinfo {year} {2018})}\BibitemShut {NoStop}%
\bibitem [{\citenamefont {Hyman}\ \emph {et~al.}(2014)\citenamefont {Hyman},
  \citenamefont {Weber},\ and\ \citenamefont {J\"{u}licher}}]{hyman2014}%
  \BibitemOpen
  \bibfield  {author} {\bibinfo {author} {\bibfnamefont {A.~A.}\ \bibnamefont
  {Hyman}}, \bibinfo {author} {\bibfnamefont {C.~A.}\ \bibnamefont {Weber}},\
  and\ \bibinfo {author} {\bibfnamefont {F.}~\bibnamefont {J\"{u}licher}},\
  }\bibfield  {title} {\bibinfo {title} {Liquid-liquid phase separation in
  biology},\ }\href {https://doi.org/10.1146/annurev-cellbio-100913-013325}
  {\bibfield  {journal} {\bibinfo  {journal} {Annu. Rev. Cell Dev. Biol.}\
  }\textbf {\bibinfo {volume} {30}},\ \bibinfo {pages} {39} (\bibinfo {year}
  {2014})}\BibitemShut {NoStop}%
\bibitem [{\citenamefont {Berry}\ \emph {et~al.}(2018)\citenamefont {Berry},
  \citenamefont {Brangwynne},\ and\ \citenamefont {Haataja}}]{Berry_2018}%
  \BibitemOpen
  \bibfield  {author} {\bibinfo {author} {\bibfnamefont {J.}~\bibnamefont
  {Berry}}, \bibinfo {author} {\bibfnamefont {C.~P.}\ \bibnamefont
  {Brangwynne}},\ and\ \bibinfo {author} {\bibfnamefont {M.}~\bibnamefont
  {Haataja}},\ }\bibfield  {title} {\bibinfo {title} {Physical principles of
  intracellular organization via active and passive phase transitions},\ }\href
  {https://doi.org/10.1088/1361-6633/aaa61e} {\bibfield  {journal} {\bibinfo
  {journal} {Rep. Prog. Phys.}\ }\textbf {\bibinfo {volume} {81}},\ \bibinfo
  {pages} {046601} (\bibinfo {year} {2018})}\BibitemShut {NoStop}%
\bibitem [{\citenamefont {Siekevitz}(1957)}]{siekevitz1957powerhouse}%
  \BibitemOpen
  \bibfield  {author} {\bibinfo {author} {\bibfnamefont {P.}~\bibnamefont
  {Siekevitz}},\ }\bibfield  {title} {\bibinfo {title} {Powerhouse of the
  cell},\ }\href {https://www.jstor.org/stable/24940890} {\bibfield  {journal}
  {\bibinfo  {journal} {Sci. Am.}\ }\textbf {\bibinfo {volume} {197}},\
  \bibinfo {pages} {131} (\bibinfo {year} {1957})}\BibitemShut {NoStop}%
\bibitem [{\citenamefont {Glover}\ \emph {et~al.}(1993)\citenamefont {Glover},
  \citenamefont {Gonzalez},\ and\ \citenamefont {Raff}}]{glover1993centrosome}%
  \BibitemOpen
  \bibfield  {author} {\bibinfo {author} {\bibfnamefont {D.~M.}\ \bibnamefont
  {Glover}}, \bibinfo {author} {\bibfnamefont {C.}~\bibnamefont {Gonzalez}},\
  and\ \bibinfo {author} {\bibfnamefont {J.~W.}\ \bibnamefont {Raff}},\
  }\bibfield  {title} {\bibinfo {title} {The centrosome},\ }\href
  {https://www.jstor.org/stable/24941512} {\bibfield  {journal} {\bibinfo
  {journal} {Sci. Am.}\ }\textbf {\bibinfo {volume} {268}},\ \bibinfo {pages}
  {62} (\bibinfo {year} {1993})}\BibitemShut {NoStop}%
\bibitem [{\citenamefont {Brangwynne}\ \emph {et~al.}(2009)\citenamefont
  {Brangwynne}, \citenamefont {Eckmann}, \citenamefont {Courson}, \citenamefont
  {Rybarska}, \citenamefont {Hoege}, \citenamefont {Gharakhani}, \citenamefont
  {Jülicher},\ and\ \citenamefont {Hyman}}]{brangwynne2009germline}%
  \BibitemOpen
  \bibfield  {author} {\bibinfo {author} {\bibfnamefont {C.~P.}\ \bibnamefont
  {Brangwynne}}, \bibinfo {author} {\bibfnamefont {C.~R.}\ \bibnamefont
  {Eckmann}}, \bibinfo {author} {\bibfnamefont {D.~S.}\ \bibnamefont
  {Courson}}, \bibinfo {author} {\bibfnamefont {A.}~\bibnamefont {Rybarska}},
  \bibinfo {author} {\bibfnamefont {C.}~\bibnamefont {Hoege}}, \bibinfo
  {author} {\bibfnamefont {J.}~\bibnamefont {Gharakhani}}, \bibinfo {author}
  {\bibfnamefont {F.}~\bibnamefont {Jülicher}},\ and\ \bibinfo {author}
  {\bibfnamefont {A.~A.}\ \bibnamefont {Hyman}},\ }\bibfield  {title} {\bibinfo
  {title} {Germline p granules are liquid droplets that localize by controlled
  dissolution/condensation},\ }\href {https://doi.org/10.1126/science.1172046}
  {\bibfield  {journal} {\bibinfo  {journal} {Science}\ }\textbf {\bibinfo
  {volume} {324}},\ \bibinfo {pages} {1729} (\bibinfo {year}
  {2009})}\BibitemShut {NoStop}%
\bibitem [{\citenamefont {Jawerth}\ \emph {et~al.}(2020)\citenamefont
  {Jawerth}, \citenamefont {Fischer-Friedrich}, \citenamefont {Saha},
  \citenamefont {Wang}, \citenamefont {Franzmann}, \citenamefont {Zhang},
  \citenamefont {Sachweh}, \citenamefont {Ruer}, \citenamefont {Ijavi},
  \citenamefont {Saha}, \citenamefont {Mahamid}, \citenamefont {Hyman},\ and\
  \citenamefont {Jülicher}}]{Jawerth2020}%
  \BibitemOpen
  \bibfield  {author} {\bibinfo {author} {\bibfnamefont {L.}~\bibnamefont
  {Jawerth}}, \bibinfo {author} {\bibfnamefont {E.}~\bibnamefont
  {Fischer-Friedrich}}, \bibinfo {author} {\bibfnamefont {S.}~\bibnamefont
  {Saha}}, \bibinfo {author} {\bibfnamefont {J.}~\bibnamefont {Wang}}, \bibinfo
  {author} {\bibfnamefont {T.}~\bibnamefont {Franzmann}}, \bibinfo {author}
  {\bibfnamefont {X.}~\bibnamefont {Zhang}}, \bibinfo {author} {\bibfnamefont
  {J.}~\bibnamefont {Sachweh}}, \bibinfo {author} {\bibfnamefont
  {M.}~\bibnamefont {Ruer}}, \bibinfo {author} {\bibfnamefont {M.}~\bibnamefont
  {Ijavi}}, \bibinfo {author} {\bibfnamefont {S.}~\bibnamefont {Saha}},
  \bibinfo {author} {\bibfnamefont {J.}~\bibnamefont {Mahamid}}, \bibinfo
  {author} {\bibfnamefont {A.~A.}\ \bibnamefont {Hyman}},\ and\ \bibinfo
  {author} {\bibfnamefont {F.}~\bibnamefont {Jülicher}},\ }\bibfield  {title}
  {\bibinfo {title} {Protein condensates as aging maxwell fluids},\ }\href
  {https://doi.org/10.1126/science.aaw4951} {\bibfield  {journal} {\bibinfo
  {journal} {Science}\ }\textbf {\bibinfo {volume} {370}},\ \bibinfo {pages}
  {1317} (\bibinfo {year} {2020})}\BibitemShut {NoStop}%
\bibitem [{\citenamefont {Wang}\ \emph {et~al.}(2021)\citenamefont {Wang},
  \citenamefont {Kelley}, \citenamefont {Milovanovic}, \citenamefont
  {Schuster},\ and\ \citenamefont {Shi}}]{Wang2021}%
  \BibitemOpen
  \bibfield  {author} {\bibinfo {author} {\bibfnamefont {H.}~\bibnamefont
  {Wang}}, \bibinfo {author} {\bibfnamefont {F.~M.}\ \bibnamefont {Kelley}},
  \bibinfo {author} {\bibfnamefont {D.}~\bibnamefont {Milovanovic}}, \bibinfo
  {author} {\bibfnamefont {B.~S.}\ \bibnamefont {Schuster}},\ and\ \bibinfo
  {author} {\bibfnamefont {Z.}~\bibnamefont {Shi}},\ }\bibfield  {title}
  {\bibinfo {title} {Surface tension and viscosity of protein condensates
  quantified by micropipette aspiration},\ }\href
  {https://doi.org/https://doi.org/10.1016/j.bpr.2021.100011} {\bibfield
  {journal} {\bibinfo  {journal} {Biophys. Rep.}\ }\textbf {\bibinfo {volume}
  {1}},\ \bibinfo {pages} {100011} (\bibinfo {year} {2021})}\BibitemShut
  {NoStop}%
\bibitem [{\citenamefont {Zwicker}(2022)}]{zwicker_reaction}%
  \BibitemOpen
  \bibfield  {author} {\bibinfo {author} {\bibfnamefont {D.}~\bibnamefont
  {Zwicker}},\ }\bibfield  {title} {\bibinfo {title} {The intertwined physics
  of active chemical reactions and phase separation},\ }\href
  {https://doi.org/https://doi.org/10.1016/j.cocis.2022.101606} {\bibfield
  {journal} {\bibinfo  {journal} {Curr. Opin. Colloid Interface Sci.}\ }\textbf
  {\bibinfo {volume} {61}},\ \bibinfo {pages} {101606} (\bibinfo {year}
  {2022})}\BibitemShut {NoStop}%
\bibitem [{\citenamefont {Datt}\ \emph {et~al.}(2015)\citenamefont {Datt},
  \citenamefont {Thampi},\ and\ \citenamefont {Govindarajan}}]{datt_2015}%
  \BibitemOpen
  \bibfield  {author} {\bibinfo {author} {\bibfnamefont {C.}~\bibnamefont
  {Datt}}, \bibinfo {author} {\bibfnamefont {S.~P.}\ \bibnamefont {Thampi}},\
  and\ \bibinfo {author} {\bibfnamefont {R.}~\bibnamefont {Govindarajan}},\
  }\bibfield  {title} {\bibinfo {title} {Morphological evolution of domains in
  spinodal decomposition},\ }\href {https://doi.org/10.1103/PhysRevE.91.010101}
  {\bibfield  {journal} {\bibinfo  {journal} {Phys. Rev. E}\ }\textbf {\bibinfo
  {volume} {91}},\ \bibinfo {pages} {010101} (\bibinfo {year}
  {2015})}\BibitemShut {NoStop}%
\bibitem [{\citenamefont {Glotzer}\ \emph {et~al.}(1994)\citenamefont
  {Glotzer}, \citenamefont {Stauffer},\ and\ \citenamefont
  {Jan}}]{Glotzer_original}%
  \BibitemOpen
  \bibfield  {author} {\bibinfo {author} {\bibfnamefont {S.~C.}\ \bibnamefont
  {Glotzer}}, \bibinfo {author} {\bibfnamefont {D.}~\bibnamefont {Stauffer}},\
  and\ \bibinfo {author} {\bibfnamefont {N.}~\bibnamefont {Jan}},\ }\bibfield
  {title} {\bibinfo {title} {Monte {C}arlo simulations of phase separation in
  chemically reactive binary mixtures},\ }\href
  {https://doi.org/10.1103/PhysRevLett.72.4109} {\bibfield  {journal} {\bibinfo
   {journal} {Phys. Rev. Lett.}\ }\textbf {\bibinfo {volume} {72}},\ \bibinfo
  {pages} {4109} (\bibinfo {year} {1994})}\BibitemShut {NoStop}%
\bibitem [{\citenamefont {Lefever}\ \emph {et~al.}(1995)\citenamefont
  {Lefever}, \citenamefont {Carati},\ and\ \citenamefont
  {Hassani}}]{Lefever_comment}%
  \BibitemOpen
  \bibfield  {author} {\bibinfo {author} {\bibfnamefont {R.}~\bibnamefont
  {Lefever}}, \bibinfo {author} {\bibfnamefont {D.}~\bibnamefont {Carati}},\
  and\ \bibinfo {author} {\bibfnamefont {N.}~\bibnamefont {Hassani}},\
  }\bibfield  {title} {\bibinfo {title} {Comment on ``{M}onte {C}arlo
  simulations of phase separation in chemically reactive binary mixtures''},\
  }\href {https://doi.org/10.1103/PhysRevLett.75.1674} {\bibfield  {journal}
  {\bibinfo  {journal} {Phys. Rev. Lett.}\ }\textbf {\bibinfo {volume} {75}},\
  \bibinfo {pages} {1674} (\bibinfo {year} {1995})}\BibitemShut {NoStop}%
\bibitem [{\citenamefont {Glotzer}\ \emph
  {et~al.}(1995{\natexlab{a}})\citenamefont {Glotzer}, \citenamefont
  {Stauffer},\ and\ \citenamefont {Jan}}]{Glotzer_reply}%
  \BibitemOpen
  \bibfield  {author} {\bibinfo {author} {\bibfnamefont {S.~C.}\ \bibnamefont
  {Glotzer}}, \bibinfo {author} {\bibfnamefont {D.}~\bibnamefont {Stauffer}},\
  and\ \bibinfo {author} {\bibfnamefont {N.}~\bibnamefont {Jan}},\ }\bibfield
  {title} {\bibinfo {title} {{G}lotzer, {S}tauffer, and {J}an reply:},\ }\href
  {https://doi.org/10.1103/PhysRevLett.75.1675} {\bibfield  {journal} {\bibinfo
   {journal} {Phys. Rev. Lett.}\ }\textbf {\bibinfo {volume} {75}},\ \bibinfo
  {pages} {1675} (\bibinfo {year} {1995}{\natexlab{a}})}\BibitemShut {NoStop}%
\bibitem [{\citenamefont {te~Vrugt}\ \emph {et~al.}(2025)\citenamefont
  {te~Vrugt}, \citenamefont {Liebchen},\ and\ \citenamefont
  {Cates}}]{vrugt2025whatexactlyisactivematter}%
  \BibitemOpen
  \bibfield  {author} {\bibinfo {author} {\bibfnamefont {M.}~\bibnamefont
  {te~Vrugt}}, \bibinfo {author} {\bibfnamefont {B.}~\bibnamefont {Liebchen}},\
  and\ \bibinfo {author} {\bibfnamefont {M.~E.}\ \bibnamefont {Cates}},\ }\href
  {https://arxiv.org/abs/2507.21621} {\bibinfo {title} {What exactly is 'active
  matter'?}} (\bibinfo {year} {2025}),\ \Eprint
  {https://arxiv.org/abs/2507.21621} {arXiv:2507.21621 [cond-mat.soft]}
  \BibitemShut {NoStop}%
\bibitem [{\citenamefont {Leibler}(1980)}]{Leibler1980}%
  \BibitemOpen
  \bibfield  {author} {\bibinfo {author} {\bibfnamefont {L.}~\bibnamefont
  {Leibler}},\ }\bibfield  {title} {\bibinfo {title} {Theory of microphase
  separation in block copolymers},\ }\href
  {https://doi.org/10.1021/ma60078a047} {\bibfield  {journal} {\bibinfo
  {journal} {Macromolecules}\ }\textbf {\bibinfo {volume} {13}},\ \bibinfo
  {pages} {1602} (\bibinfo {year} {1980})}\BibitemShut {NoStop}%
\bibitem [{\citenamefont {Glotzer}\ and\ \citenamefont
  {Coniglio}(1994)}]{glotzer1994_block}%
  \BibitemOpen
  \bibfield  {author} {\bibinfo {author} {\bibfnamefont {S.~C.}\ \bibnamefont
  {Glotzer}}\ and\ \bibinfo {author} {\bibfnamefont {A.}~\bibnamefont
  {Coniglio}},\ }\bibfield  {title} {\bibinfo {title} {Self-consistent solution
  of phase separation with competing interactions},\ }\href
  {https://doi.org/10.1103/PhysRevE.50.4241} {\bibfield  {journal} {\bibinfo
  {journal} {Phys. Rev. E}\ }\textbf {\bibinfo {volume} {50}},\ \bibinfo
  {pages} {4241} (\bibinfo {year} {1994})}\BibitemShut {NoStop}%
\bibitem [{\citenamefont {Sagui}\ and\ \citenamefont
  {Desai}(1994)}]{Sagui1994}%
  \BibitemOpen
  \bibfield  {author} {\bibinfo {author} {\bibfnamefont {C.}~\bibnamefont
  {Sagui}}\ and\ \bibinfo {author} {\bibfnamefont {R.~C.}\ \bibnamefont
  {Desai}},\ }\bibfield  {title} {\bibinfo {title} {Kinetics of phase
  separation in two-dimensional systems with competing interactions},\ }\href
  {https://doi.org/10.1103/PhysRevE.49.2225} {\bibfield  {journal} {\bibinfo
  {journal} {Phys. Rev. E}\ }\textbf {\bibinfo {volume} {49}},\ \bibinfo
  {pages} {2225} (\bibinfo {year} {1994})}\BibitemShut {NoStop}%
\bibitem [{\citenamefont {Goldstein}\ \emph {et~al.}(1991)\citenamefont
  {Goldstein}, \citenamefont {Gunaratne}, \citenamefont {Gil},\ and\
  \citenamefont {Coullet}}]{goldstein1991}%
  \BibitemOpen
  \bibfield  {author} {\bibinfo {author} {\bibfnamefont {R.~E.}\ \bibnamefont
  {Goldstein}}, \bibinfo {author} {\bibfnamefont {G.~H.}\ \bibnamefont
  {Gunaratne}}, \bibinfo {author} {\bibfnamefont {L.}~\bibnamefont {Gil}},\
  and\ \bibinfo {author} {\bibfnamefont {P.}~\bibnamefont {Coullet}},\
  }\bibfield  {title} {\bibinfo {title} {Hydrodynamic and interfacial patterns
  with broken space-time symmetry},\ }\href
  {https://doi.org/10.1103/PhysRevA.43.6700} {\bibfield  {journal} {\bibinfo
  {journal} {Phys. Rev. A}\ }\textbf {\bibinfo {volume} {43}},\ \bibinfo
  {pages} {6700} (\bibinfo {year} {1991})}\BibitemShut {NoStop}%
\bibitem [{\citenamefont {Hohenberg}\ and\ \citenamefont
  {Halperin}(1977)}]{hohenberg1977}%
  \BibitemOpen
  \bibfield  {author} {\bibinfo {author} {\bibfnamefont {P.~C.}\ \bibnamefont
  {Hohenberg}}\ and\ \bibinfo {author} {\bibfnamefont {B.~I.}\ \bibnamefont
  {Halperin}},\ }\bibfield  {title} {\bibinfo {title} {Theory of dynamic
  critical phenomena},\ }\href {https://doi.org/10.1103/RevModPhys.49.435}
  {\bibfield  {journal} {\bibinfo  {journal} {Rev. Mod. Phys.}\ }\textbf
  {\bibinfo {volume} {49}},\ \bibinfo {pages} {435} (\bibinfo {year}
  {1977})}\BibitemShut {NoStop}%
\bibitem [{\citenamefont {Huo}\ \emph {et~al.}(2003)\citenamefont {Huo},
  \citenamefont {Jiang}, \citenamefont {Zhang},\ and\ \citenamefont
  {Yang}}]{Huo2003}%
  \BibitemOpen
  \bibfield  {author} {\bibinfo {author} {\bibfnamefont {Y.}~\bibnamefont
  {Huo}}, \bibinfo {author} {\bibfnamefont {X.}~\bibnamefont {Jiang}}, \bibinfo
  {author} {\bibfnamefont {H.}~\bibnamefont {Zhang}},\ and\ \bibinfo {author}
  {\bibfnamefont {Y.}~\bibnamefont {Yang}},\ }\bibfield  {title} {\bibinfo
  {title} {{Hydrodynamic effects on phase separation of binary mixtures with
  reversible chemical reaction}},\ }\href {https://doi.org/10.1063/1.1571511}
  {\bibfield  {journal} {\bibinfo  {journal} {J. Chem. Phys.}\ }\textbf
  {\bibinfo {volume} {118}},\ \bibinfo {pages} {9830} (\bibinfo {year}
  {2003})}\BibitemShut {NoStop}%
\bibitem [{\citenamefont {Ascher}\ \emph {et~al.}(1997)\citenamefont {Ascher},
  \citenamefont {Ruuth},\ and\ \citenamefont {Spiteri}}]{ascher1997}%
  \BibitemOpen
  \bibfield  {author} {\bibinfo {author} {\bibfnamefont {U.~M.}\ \bibnamefont
  {Ascher}}, \bibinfo {author} {\bibfnamefont {S.~J.}\ \bibnamefont {Ruuth}},\
  and\ \bibinfo {author} {\bibfnamefont {R.~J.}\ \bibnamefont {Spiteri}},\
  }\bibfield  {title} {\bibinfo {title} {Implicit-explicit runge-kutta methods
  for time-dependent partial differential equations},\ }\href
  {https://doi.org/https://doi.org/10.1016/S0168-9274(97)00056-1} {\bibfield
  {journal} {\bibinfo  {journal} {Appl. Num. Math.}\ }\textbf {\bibinfo
  {volume} {25}},\ \bibinfo {pages} {151} (\bibinfo {year} {1997})}\BibitemShut
  {NoStop}%
\bibitem [{\citenamefont {Padhan}\ \emph {et~al.}(2024)\citenamefont {Padhan},
  \citenamefont {Kiran},\ and\ \citenamefont {Pandit}}]{pandit2024}%
  \BibitemOpen
  \bibfield  {author} {\bibinfo {author} {\bibfnamefont {N.~B.}\ \bibnamefont
  {Padhan}}, \bibinfo {author} {\bibfnamefont {K.~V.}\ \bibnamefont {Kiran}},\
  and\ \bibinfo {author} {\bibfnamefont {R.}~\bibnamefont {Pandit}},\
  }\bibfield  {title} {\bibinfo {title} {Novel turbulence and coarsening arrest
  in active-scalar fluids},\ }\href {https://doi.org/10.1039/D4SM00163J}
  {\bibfield  {journal} {\bibinfo  {journal} {Soft Matter}\ }\textbf {\bibinfo
  {volume} {20}},\ \bibinfo {pages} {3620} (\bibinfo {year}
  {2024})}\BibitemShut {NoStop}%
\bibitem [{\citenamefont {Derevyanko}(2008)}]{dealias_paper}%
  \BibitemOpen
  \bibfield  {author} {\bibinfo {author} {\bibfnamefont {S.}~\bibnamefont
  {Derevyanko}},\ }\bibfield  {title} {\bibinfo {title} {The $(n+1)/2$ rule for
  dealiasing in the split-step fourier methods for $n$-wave interactions},\
  }\href {https://doi.org/10.1109/LPT.2008.2005420} {\bibfield  {journal}
  {\bibinfo  {journal} {IEEE Photon. Technol. Lett.}\ }\textbf {\bibinfo
  {volume} {20}},\ \bibinfo {pages} {1929} (\bibinfo {year}
  {2008})}\BibitemShut {NoStop}%
\bibitem [{\citenamefont {Choksi}\ \emph {et~al.}(2011)\citenamefont {Choksi},
  \citenamefont {Maras},\ and\ \citenamefont {Williams}}]{rustum2011}%
  \BibitemOpen
  \bibfield  {author} {\bibinfo {author} {\bibfnamefont {R.}~\bibnamefont
  {Choksi}}, \bibinfo {author} {\bibfnamefont {M.}~\bibnamefont {Maras}},\ and\
  \bibinfo {author} {\bibfnamefont {J.~F.}\ \bibnamefont {Williams}},\
  }\bibfield  {title} {\bibinfo {title} {2d phase diagram for minimizers of a
  cahn–hilliard functional with long-range interactions},\ }\href
  {https://doi.org/10.1137/100784497} {\bibfield  {journal} {\bibinfo
  {journal} {SIAM J. Appl. Dyn. Syst.}\ }\textbf {\bibinfo {volume} {10}},\
  \bibinfo {pages} {1344} (\bibinfo {year} {2011})}\BibitemShut {NoStop}%
\bibitem [{\citenamefont {Glotzer}\ \emph
  {et~al.}(1995{\natexlab{b}})\citenamefont {Glotzer}, \citenamefont
  {Di~Marzio},\ and\ \citenamefont {Muthukumar}}]{glotzer1995}%
  \BibitemOpen
  \bibfield  {author} {\bibinfo {author} {\bibfnamefont {S.~C.}\ \bibnamefont
  {Glotzer}}, \bibinfo {author} {\bibfnamefont {E.~A.}\ \bibnamefont
  {Di~Marzio}},\ and\ \bibinfo {author} {\bibfnamefont {M.}~\bibnamefont
  {Muthukumar}},\ }\bibfield  {title} {\bibinfo {title} {Reaction-controlled
  morphology of phase-separating mixtures},\ }\href
  {https://doi.org/10.1103/PhysRevLett.74.2034} {\bibfield  {journal} {\bibinfo
   {journal} {Phys. Rev. Lett.}\ }\textbf {\bibinfo {volume} {74}},\ \bibinfo
  {pages} {2034} (\bibinfo {year} {1995}{\natexlab{b}})}\BibitemShut {NoStop}%
\bibitem [{\citenamefont {Liu}\ and\ \citenamefont
  {Goldenfeld}(1989)}]{Liu1989}%
  \BibitemOpen
  \bibfield  {author} {\bibinfo {author} {\bibfnamefont {F.}~\bibnamefont
  {Liu}}\ and\ \bibinfo {author} {\bibfnamefont {N.}~\bibnamefont
  {Goldenfeld}},\ }\bibfield  {title} {\bibinfo {title} {Dynamics of phase
  separation in block copolymer melts},\ }\href
  {https://doi.org/10.1103/PhysRevA.39.4805} {\bibfield  {journal} {\bibinfo
  {journal} {Phys. Rev. A}\ }\textbf {\bibinfo {volume} {39}},\ \bibinfo
  {pages} {4805} (\bibinfo {year} {1989})}\BibitemShut {NoStop}%
\bibitem [{\citenamefont {Seyboldt}\ and\ \citenamefont
  {Jülicher}(2018)}]{Seyboldt_2018}%
  \BibitemOpen
  \bibfield  {author} {\bibinfo {author} {\bibfnamefont {R.}~\bibnamefont
  {Seyboldt}}\ and\ \bibinfo {author} {\bibfnamefont {F.}~\bibnamefont
  {Jülicher}},\ }\bibfield  {title} {\bibinfo {title} {Role of hydrodynamic
  flows in chemically driven droplet division},\ }\href
  {https://doi.org/10.1088/1367-2630/aae735} {\bibfield  {journal} {\bibinfo
  {journal} {New J. Phys.}\ }\textbf {\bibinfo {volume} {20}},\ \bibinfo
  {pages} {105010} (\bibinfo {year} {2018})}\BibitemShut {NoStop}%
\bibitem [{\citenamefont {Karimi}\ \emph {et~al.}(2011)\citenamefont {Karimi},
  \citenamefont {Huang},\ and\ \citenamefont {Paul}}]{karimi2011}%
  \BibitemOpen
  \bibfield  {author} {\bibinfo {author} {\bibfnamefont {A.}~\bibnamefont
  {Karimi}}, \bibinfo {author} {\bibfnamefont {Z.-F.}\ \bibnamefont {Huang}},\
  and\ \bibinfo {author} {\bibfnamefont {M.~R.}\ \bibnamefont {Paul}},\
  }\bibfield  {title} {\bibinfo {title} {Exploring spiral defect chaos in
  generalized {S}wift-{H}ohenberg models with mean flow},\ }\href
  {https://doi.org/10.1103/PhysRevE.84.046215} {\bibfield  {journal} {\bibinfo
  {journal} {Phys. Rev. E}\ }\textbf {\bibinfo {volume} {84}},\ \bibinfo
  {pages} {046215} (\bibinfo {year} {2011})}\BibitemShut {NoStop}%
\bibitem [{\citenamefont {Chiam}\ \emph
  {et~al.}(2003{\natexlab{a}})\citenamefont {Chiam}, \citenamefont {Paul},
  \citenamefont {Cross},\ and\ \citenamefont {Greenside}}]{Chaim2003}%
  \BibitemOpen
  \bibfield  {author} {\bibinfo {author} {\bibfnamefont {K.-H.}\ \bibnamefont
  {Chiam}}, \bibinfo {author} {\bibfnamefont {M.~R.}\ \bibnamefont {Paul}},
  \bibinfo {author} {\bibfnamefont {M.~C.}\ \bibnamefont {Cross}},\ and\
  \bibinfo {author} {\bibfnamefont {H.~S.}\ \bibnamefont {Greenside}},\
  }\bibfield  {title} {\bibinfo {title} {Mean flow and spiral defect chaos in
  {R}ayleigh-{B}\'enard convection},\ }\href
  {https://doi.org/10.1103/PhysRevE.67.056206} {\bibfield  {journal} {\bibinfo
  {journal} {Phys. Rev. E}\ }\textbf {\bibinfo {volume} {67}},\ \bibinfo
  {pages} {056206} (\bibinfo {year} {2003}{\natexlab{a}})}\BibitemShut
  {NoStop}%
\bibitem [{\citenamefont {Furtado}\ and\ \citenamefont
  {Yeomans}(2006)}]{furtado2006}%
  \BibitemOpen
  \bibfield  {author} {\bibinfo {author} {\bibfnamefont {K.}~\bibnamefont
  {Furtado}}\ and\ \bibinfo {author} {\bibfnamefont {J.~M.}\ \bibnamefont
  {Yeomans}},\ }\bibfield  {title} {\bibinfo {title} {Lattice {B}oltzmann
  simulations of phase separation in chemically reactive binary fluids},\
  }\href {https://doi.org/10.1103/PhysRevE.73.066124} {\bibfield  {journal}
  {\bibinfo  {journal} {Phys. Rev. E}\ }\textbf {\bibinfo {volume} {73}},\
  \bibinfo {pages} {066124} (\bibinfo {year} {2006})}\BibitemShut {NoStop}%
\bibitem [{\citenamefont {Benettin}\ \emph {et~al.}(1976)\citenamefont
  {Benettin}, \citenamefont {Galgani},\ and\ \citenamefont
  {Strelcyn}}]{benettin1976}%
  \BibitemOpen
  \bibfield  {author} {\bibinfo {author} {\bibfnamefont {G.}~\bibnamefont
  {Benettin}}, \bibinfo {author} {\bibfnamefont {L.}~\bibnamefont {Galgani}},\
  and\ \bibinfo {author} {\bibfnamefont {J.-M.}\ \bibnamefont {Strelcyn}},\
  }\bibfield  {title} {\bibinfo {title} {Kolmogorov entropy and numerical
  experiments},\ }\href {https://doi.org/10.1103/PhysRevA.14.2338} {\bibfield
  {journal} {\bibinfo  {journal} {Phys. Rev. A}\ }\textbf {\bibinfo {volume}
  {14}},\ \bibinfo {pages} {2338} (\bibinfo {year} {1976})}\BibitemShut
  {NoStop}%
\bibitem [{\citenamefont {Tanaka}(1995)}]{tanaka1995}%
  \BibitemOpen
  \bibfield  {author} {\bibinfo {author} {\bibfnamefont {H.}~\bibnamefont
  {Tanaka}},\ }\bibfield  {title} {\bibinfo {title} {Hydrodynamic interface
  quench effects on spinodal decomposition for symmetric binary fluid
  mixtures},\ }\href {https://doi.org/10.1103/PhysRevE.51.1313} {\bibfield
  {journal} {\bibinfo  {journal} {Phys. Rev. E}\ }\textbf {\bibinfo {volume}
  {51}},\ \bibinfo {pages} {1313} (\bibinfo {year} {1995})}\BibitemShut
  {NoStop}%
\bibitem [{\citenamefont {Hoyle}(2006)}]{hoyle2006pattern}%
  \BibitemOpen
  \bibfield  {author} {\bibinfo {author} {\bibfnamefont {R.~B.}\ \bibnamefont
  {Hoyle}},\ }\href@noop {} {\emph {\bibinfo {title} {Pattern formation: an
  introduction to methods}}}\ (\bibinfo  {publisher} {Cambridge University
  Press},\ \bibinfo {year} {2006})\BibitemShut {NoStop}%
\bibitem [{\citenamefont {Shiwa}(1997)}]{shiwa1997amplitude}%
  \BibitemOpen
  \bibfield  {author} {\bibinfo {author} {\bibfnamefont {Y.}~\bibnamefont
  {Shiwa}},\ }\bibfield  {title} {\bibinfo {title} {The amplitude and
  phase-diffusion equations for lamellar patterns in block copolymers},\ }\href
  {https://doi.org/https://doi.org/10.1016/S0375-9601(97)00128-X} {\bibfield
  {journal} {\bibinfo  {journal} {Phys. Lett. A}\ }\textbf {\bibinfo {volume}
  {228}},\ \bibinfo {pages} {279} (\bibinfo {year} {1997})}\BibitemShut
  {NoStop}%
\bibitem [{\citenamefont {Gunaratne}\ \emph {et~al.}(1994)\citenamefont
  {Gunaratne}, \citenamefont {Ouyang},\ and\ \citenamefont
  {Swinney}}]{gunaratne1994pattern}%
  \BibitemOpen
  \bibfield  {author} {\bibinfo {author} {\bibfnamefont {G.~H.}\ \bibnamefont
  {Gunaratne}}, \bibinfo {author} {\bibfnamefont {Q.}~\bibnamefont {Ouyang}},\
  and\ \bibinfo {author} {\bibfnamefont {H.~L.}\ \bibnamefont {Swinney}},\
  }\bibfield  {title} {\bibinfo {title} {Pattern formation in the presence of
  symmetries},\ }\href {https://doi.org/10.1103/PhysRevE.50.2802} {\bibfield
  {journal} {\bibinfo  {journal} {Phys. Rev. E}\ }\textbf {\bibinfo {volume}
  {50}},\ \bibinfo {pages} {2802} (\bibinfo {year} {1994})}\BibitemShut
  {NoStop}%
\bibitem [{\citenamefont {Siggia}\ and\ \citenamefont
  {Zippelius}(1981)}]{siggia1981}%
  \BibitemOpen
  \bibfield  {author} {\bibinfo {author} {\bibfnamefont {E.~D.}\ \bibnamefont
  {Siggia}}\ and\ \bibinfo {author} {\bibfnamefont {A.}~\bibnamefont
  {Zippelius}},\ }\bibfield  {title} {\bibinfo {title} {Pattern selection in
  {R}ayleigh-{B}\'enard convection near threshold},\ }\href
  {https://doi.org/10.1103/PhysRevLett.47.835} {\bibfield  {journal} {\bibinfo
  {journal} {Phys. Rev. Lett.}\ }\textbf {\bibinfo {volume} {47}},\ \bibinfo
  {pages} {835} (\bibinfo {year} {1981})}\BibitemShut {NoStop}%
\bibitem [{\citenamefont {Matthews}(1998)}]{Matthews1998}%
  \BibitemOpen
  \bibfield  {author} {\bibinfo {author} {\bibfnamefont {P.}~\bibnamefont
  {Matthews}},\ }\bibfield  {title} {\bibinfo {title} {Hexagonal patterns in
  finite domains},\ }\href
  {https://doi.org/https://doi.org/10.1016/S0167-2789(97)00248-0} {\bibfield
  {journal} {\bibinfo  {journal} {Phys. D}\ }\textbf {\bibinfo {volume}
  {116}},\ \bibinfo {pages} {81} (\bibinfo {year} {1998})}\BibitemShut
  {NoStop}%
\bibitem [{\citenamefont {Young}\ and\ \citenamefont
  {Riecke}(2002)}]{young_mean_flow}%
  \BibitemOpen
  \bibfield  {author} {\bibinfo {author} {\bibfnamefont {Y.-N.}\ \bibnamefont
  {Young}}\ and\ \bibinfo {author} {\bibfnamefont {H.}~\bibnamefont {Riecke}},\
  }\bibfield  {title} {\bibinfo {title} {Mean flow in hexagonal convection:
  stability and nonlinear dynamics},\ }\href
  {https://doi.org/https://doi.org/10.1016/S0167-2789(01)00389-X} {\bibfield
  {journal} {\bibinfo  {journal} {Phys. D}\ }\textbf {\bibinfo {volume}
  {163}},\ \bibinfo {pages} {166} (\bibinfo {year} {2002})}\BibitemShut
  {NoStop}%
\bibitem [{\citenamefont {Newell}\ and\ \citenamefont
  {Whitehead}(1969)}]{Newell_Whitehead_1969}%
  \BibitemOpen
  \bibfield  {author} {\bibinfo {author} {\bibfnamefont {A.~C.}\ \bibnamefont
  {Newell}}\ and\ \bibinfo {author} {\bibfnamefont {J.~A.}\ \bibnamefont
  {Whitehead}},\ }\bibfield  {title} {\bibinfo {title} {Finite bandwidth,
  finite amplitude convection},\ }\href
  {https://doi.org/10.1017/S0022112069000176} {\bibfield  {journal} {\bibinfo
  {journal} {J. Fluid Mech.}\ }\textbf {\bibinfo {volume} {38}},\ \bibinfo
  {pages} {279–303} (\bibinfo {year} {1969})}\BibitemShut {NoStop}%
\bibitem [{\citenamefont {Segel}(1969)}]{Segel_1969}%
  \BibitemOpen
  \bibfield  {author} {\bibinfo {author} {\bibfnamefont {L.~A.}\ \bibnamefont
  {Segel}},\ }\bibfield  {title} {\bibinfo {title} {Distant side-walls cause
  slow amplitude modulation of cellular convection},\ }\href
  {https://doi.org/10.1017/S0022112069000127} {\bibfield  {journal} {\bibinfo
  {journal} {J. Fluid Mech.}\ }\textbf {\bibinfo {volume} {38}},\ \bibinfo
  {pages} {203–224} (\bibinfo {year} {1969})}\BibitemShut {NoStop}%
\bibitem [{\citenamefont {Chiam}\ \emph
  {et~al.}(2003{\natexlab{b}})\citenamefont {Chiam}, \citenamefont {Paul},
  \citenamefont {Cross},\ and\ \citenamefont {Greenside}}]{chiam2003}%
  \BibitemOpen
  \bibfield  {author} {\bibinfo {author} {\bibfnamefont {K.-H.}\ \bibnamefont
  {Chiam}}, \bibinfo {author} {\bibfnamefont {M.~R.}\ \bibnamefont {Paul}},
  \bibinfo {author} {\bibfnamefont {M.~C.}\ \bibnamefont {Cross}},\ and\
  \bibinfo {author} {\bibfnamefont {H.~S.}\ \bibnamefont {Greenside}},\
  }\bibfield  {title} {\bibinfo {title} {Mean flow and spiral defect chaos in
  {R}ayleigh-{B}\'enard convection},\ }\href
  {https://doi.org/10.1103/PhysRevE.67.056206} {\bibfield  {journal} {\bibinfo
  {journal} {Phys. Rev. E}\ }\textbf {\bibinfo {volume} {67}},\ \bibinfo
  {pages} {056206} (\bibinfo {year} {2003}{\natexlab{b}})}\BibitemShut
  {NoStop}%
\bibitem [{\citenamefont {Bodenschatz}\ \emph {et~al.}(2000)\citenamefont
  {Bodenschatz}, \citenamefont {Pesch},\ and\ \citenamefont
  {Ahlers}}]{Bodenschatz2000}%
  \BibitemOpen
  \bibfield  {author} {\bibinfo {author} {\bibfnamefont {E.}~\bibnamefont
  {Bodenschatz}}, \bibinfo {author} {\bibfnamefont {W.}~\bibnamefont {Pesch}},\
  and\ \bibinfo {author} {\bibfnamefont {G.}~\bibnamefont {Ahlers}},\
  }\bibfield  {title} {\bibinfo {title} {Recent developments in
  {R}ayleigh-{B}énard convection},\ }\href
  {https://doi.org/https://doi.org/10.1146/annurev.fluid.32.1.709} {\bibfield
  {journal} {\bibinfo  {journal} {Annu. Rev. Fluid Mech.}\ }\textbf {\bibinfo
  {volume} {32}},\ \bibinfo {pages} {709} (\bibinfo {year} {2000})}\BibitemShut
  {NoStop}%
\bibitem [{\citenamefont {Shiwa}(2005)}]{shiwa2005a}%
  \BibitemOpen
  \bibfield  {author} {\bibinfo {author} {\bibfnamefont {Y.}~\bibnamefont
  {Shiwa}},\ }\bibfield  {title} {\bibinfo {title} {Hydrodynamic coarsening in
  striped pattern formation with a conservation law},\ }\href
  {https://doi.org/10.1103/PhysRevE.72.016204} {\bibfield  {journal} {\bibinfo
  {journal} {Phys. Rev. E}\ }\textbf {\bibinfo {volume} {72}},\ \bibinfo
  {pages} {016204} (\bibinfo {year} {2005})}\BibitemShut {NoStop}%
\bibitem [{\citenamefont {Shiwa}(2006)}]{shiwa_conference}%
  \BibitemOpen
  \bibfield  {author} {\bibinfo {author} {\bibfnamefont {Y.}~\bibnamefont
  {Shiwa}},\ }\bibfield  {title} {\bibinfo {title} {Spirals, hexagons, and all
  that through hydrodynamic coupling with a zero mode},\ }\href
  {https://doi.org/10.1063/1.2204492} {\bibfield  {journal} {\bibinfo
  {journal} {AIP Conf. Proc.}\ }\textbf {\bibinfo {volume} {832}},\ \bibinfo
  {pages} {205} (\bibinfo {year} {2006})}\BibitemShut {NoStop}%
\bibitem [{\citenamefont {Ohnogi}\ and\ \citenamefont
  {Shiwa}(2011)}]{Ohnogi2011}%
  \BibitemOpen
  \bibfield  {author} {\bibinfo {author} {\bibfnamefont {H.}~\bibnamefont
  {Ohnogi}}\ and\ \bibinfo {author} {\bibfnamefont {Y.}~\bibnamefont {Shiwa}},\
  }\bibfield  {title} {\bibinfo {title} {Nucleation, growth, and coarsening of
  crystalline domains in order-order transitions between lamellar and hexagonal
  phases},\ }\href {https://doi.org/10.1103/PhysRevE.84.011611} {\bibfield
  {journal} {\bibinfo  {journal} {Phys. Rev. E}\ }\textbf {\bibinfo {volume}
  {84}},\ \bibinfo {pages} {011611} (\bibinfo {year} {2011})}\BibitemShut
  {NoStop}%
\bibitem [{\citenamefont {McLeish}\ \emph {et~al.}(2003)\citenamefont
  {McLeish}, \citenamefont {Cates}, \citenamefont {Higgins}, \citenamefont
  {Olmsted}, \citenamefont {Cates}, \citenamefont {Vollmer}, \citenamefont
  {Wagner},\ and\ \citenamefont {Vollmer}}]{cates2003}%
  \BibitemOpen
  \bibfield  {author} {\bibinfo {author} {\bibfnamefont {T.~C.~B.}\
  \bibnamefont {McLeish}}, \bibinfo {author} {\bibfnamefont {M.~E.}\
  \bibnamefont {Cates}}, \bibinfo {author} {\bibfnamefont {J.~S.}\ \bibnamefont
  {Higgins}}, \bibinfo {author} {\bibfnamefont {P.~D.}\ \bibnamefont
  {Olmsted}}, \bibinfo {author} {\bibfnamefont {M.~E.}\ \bibnamefont {Cates}},
  \bibinfo {author} {\bibfnamefont {J.}~\bibnamefont {Vollmer}}, \bibinfo
  {author} {\bibfnamefont {A.}~\bibnamefont {Wagner}},\ and\ \bibinfo {author}
  {\bibfnamefont {D.}~\bibnamefont {Vollmer}},\ }\bibfield  {title} {\bibinfo
  {title} {Phase separation in binary fluid mixtures with continuously ramped
  temperature},\ }\href {https://doi.org/10.1098/rsta.2002.1165} {\bibfield
  {journal} {\bibinfo  {journal} {Phil. Trans. R. Soc. A.}\ }\textbf {\bibinfo
  {volume} {361}},\ \bibinfo {pages} {793} (\bibinfo {year}
  {2003})}\BibitemShut {NoStop}%
\bibitem [{\citenamefont {Golovin}\ \emph {et~al.}(2001)\citenamefont
  {Golovin}, \citenamefont {Nepomnyashchy}, \citenamefont {Davis},\ and\
  \citenamefont {Zaks}}]{golovin2001}%
  \BibitemOpen
  \bibfield  {author} {\bibinfo {author} {\bibfnamefont {A.~A.}\ \bibnamefont
  {Golovin}}, \bibinfo {author} {\bibfnamefont {A.~A.}\ \bibnamefont
  {Nepomnyashchy}}, \bibinfo {author} {\bibfnamefont {S.~H.}\ \bibnamefont
  {Davis}},\ and\ \bibinfo {author} {\bibfnamefont {M.~A.}\ \bibnamefont
  {Zaks}},\ }\bibfield  {title} {\bibinfo {title} {Convective {C}ahn-{H}illiard
  models: From coarsening to roughening},\ }\href
  {https://doi.org/10.1103/PhysRevLett.86.1550} {\bibfield  {journal} {\bibinfo
   {journal} {Phys. Rev. Lett.}\ }\textbf {\bibinfo {volume} {86}},\ \bibinfo
  {pages} {1550} (\bibinfo {year} {2001})}\BibitemShut {NoStop}%
\bibitem [{\citenamefont {Golovin}\ and\ \citenamefont
  {Pismen}(2004)}]{golovin2004}%
  \BibitemOpen
  \bibfield  {author} {\bibinfo {author} {\bibfnamefont {A.~A.}\ \bibnamefont
  {Golovin}}\ and\ \bibinfo {author} {\bibfnamefont {L.~M.}\ \bibnamefont
  {Pismen}},\ }\bibfield  {title} {\bibinfo {title} {{Dynamic phase separation:
  From coarsening to turbulence via structure formation}},\ }\href
  {https://doi.org/10.1063/1.1784751} {\bibfield  {journal} {\bibinfo
  {journal} {Chaos}\ }\textbf {\bibinfo {volume} {14}},\ \bibinfo {pages} {845}
  (\bibinfo {year} {2004})}\BibitemShut {NoStop}%
\bibitem [{\citenamefont {Miike}\ \emph {et~al.}(1988)\citenamefont {Miike},
  \citenamefont {M\"uller},\ and\ \citenamefont {Hess}}]{Miike1988}%
  \BibitemOpen
  \bibfield  {author} {\bibinfo {author} {\bibfnamefont {H.}~\bibnamefont
  {Miike}}, \bibinfo {author} {\bibfnamefont {S.~C.}\ \bibnamefont
  {M\"uller}},\ and\ \bibinfo {author} {\bibfnamefont {B.}~\bibnamefont
  {Hess}},\ }\bibfield  {title} {\bibinfo {title} {Oscillatory deformation of
  chemical waves induced by surface flow},\ }\href
  {https://doi.org/10.1103/PhysRevLett.61.2109} {\bibfield  {journal} {\bibinfo
   {journal} {Phys. Rev. Lett.}\ }\textbf {\bibinfo {volume} {61}},\ \bibinfo
  {pages} {2109} (\bibinfo {year} {1988})}\BibitemShut {NoStop}%
\bibitem [{\citenamefont {Miike}\ \emph {et~al.}(1989)\citenamefont {Miike},
  \citenamefont {Müller},\ and\ \citenamefont {Hess}}]{Miike1989}%
  \BibitemOpen
  \bibfield  {author} {\bibinfo {author} {\bibfnamefont {H.}~\bibnamefont
  {Miike}}, \bibinfo {author} {\bibfnamefont {S.~C.}\ \bibnamefont {Müller}},\
  and\ \bibinfo {author} {\bibfnamefont {B.}~\bibnamefont {Hess}},\ }\bibfield
  {title} {\bibinfo {title} {Hydrodynamic flows traveling with chemical
  waves},\ }\href
  {https://doi.org/https://doi.org/10.1016/0375-9601(89)90438-6} {\bibfield
  {journal} {\bibinfo  {journal} {Phys. Lett. A}\ }\textbf {\bibinfo {volume}
  {141}},\ \bibinfo {pages} {25} (\bibinfo {year} {1989})}\BibitemShut
  {NoStop}%
\bibitem [{\citenamefont {Diewald}\ and\ \citenamefont
  {Brand}(1995)}]{Diewald1995}%
  \BibitemOpen
  \bibfield  {author} {\bibinfo {author} {\bibfnamefont {M.}~\bibnamefont
  {Diewald}}\ and\ \bibinfo {author} {\bibfnamefont {H.~R.}\ \bibnamefont
  {Brand}},\ }\bibfield  {title} {\bibinfo {title} {Chemically driven
  convection can stabilize turing patterns},\ }\href
  {https://doi.org/10.1103/PhysRevE.51.R5200} {\bibfield  {journal} {\bibinfo
  {journal} {Phys. Rev. E}\ }\textbf {\bibinfo {volume} {51}},\ \bibinfo
  {pages} {R5200} (\bibinfo {year} {1995})}\BibitemShut {NoStop}%
\bibitem [{\citenamefont {De~Wit}(2020)}]{DeWit_review}%
  \BibitemOpen
  \bibfield  {author} {\bibinfo {author} {\bibfnamefont {A.}~\bibnamefont
  {De~Wit}},\ }\bibfield  {title} {\bibinfo {title} {Chemo-hydrodynamic
  patterns and instabilities},\ }\href
  {https://doi.org/https://doi.org/10.1146/annurev-fluid-010719-060349}
  {\bibfield  {journal} {\bibinfo  {journal} {Annu. Rev. Fluid Mech.}\ }\textbf
  {\bibinfo {volume} {52}},\ \bibinfo {pages} {531} (\bibinfo {year}
  {2020})}\BibitemShut {NoStop}%
\bibitem [{\citenamefont {Huffman}\ and\ \citenamefont
  {Shum}(2023)}]{Huffman2023}%
  \BibitemOpen
  \bibfield  {author} {\bibinfo {author} {\bibfnamefont {A.}~\bibnamefont
  {Huffman}}\ and\ \bibinfo {author} {\bibfnamefont {H.}~\bibnamefont {Shum}},\
  }\bibfield  {title} {\bibinfo {title} {Boundary-bound reactions: Pattern
  formation with and without hydrodynamics},\ }\href
  {https://doi.org/10.1103/PhysRevE.108.055103} {\bibfield  {journal} {\bibinfo
   {journal} {Phys. Rev. E}\ }\textbf {\bibinfo {volume} {108}},\ \bibinfo
  {pages} {055103} (\bibinfo {year} {2023})}\BibitemShut {NoStop}%
\bibitem [{\citenamefont {Tiribocchi}\ \emph {et~al.}(2015)\citenamefont
  {Tiribocchi}, \citenamefont {Wittkowski}, \citenamefont {Marenduzzo},\ and\
  \citenamefont {Cates}}]{tiribocchi2015}%
  \BibitemOpen
  \bibfield  {author} {\bibinfo {author} {\bibfnamefont {A.}~\bibnamefont
  {Tiribocchi}}, \bibinfo {author} {\bibfnamefont {R.}~\bibnamefont
  {Wittkowski}}, \bibinfo {author} {\bibfnamefont {D.}~\bibnamefont
  {Marenduzzo}},\ and\ \bibinfo {author} {\bibfnamefont {M.~E.}\ \bibnamefont
  {Cates}},\ }\bibfield  {title} {\bibinfo {title} {Active model {H}: Scalar
  active matter in a momentum-conserving fluid},\ }\href
  {https://doi.org/10.1103/PhysRevLett.115.188302} {\bibfield  {journal}
  {\bibinfo  {journal} {Phys. Rev. Lett.}\ }\textbf {\bibinfo {volume} {115}},\
  \bibinfo {pages} {188302} (\bibinfo {year} {2015})}\BibitemShut {NoStop}%
\bibitem [{\citenamefont {Lowengrub}\ and\ \citenamefont
  {Truskinovsky}(1998)}]{lowengrub1978}%
  \BibitemOpen
  \bibfield  {author} {\bibinfo {author} {\bibfnamefont {J.}~\bibnamefont
  {Lowengrub}}\ and\ \bibinfo {author} {\bibfnamefont {L.}~\bibnamefont
  {Truskinovsky}},\ }\bibfield  {title} {\bibinfo {title}
  {Quasi–incompressible {C}ahn–{H}illiard fluids and topological
  transitions},\ }\href {https://doi.org/10.1098/rspa.1998.0273} {\bibfield
  {journal} {\bibinfo  {journal} {Proc. R. Soc. Lond. A.}\ }\textbf {\bibinfo
  {volume} {454}},\ \bibinfo {pages} {2617} (\bibinfo {year}
  {1998})}\BibitemShut {NoStop}%
\bibitem [{\citenamefont {Onuki}(1994)}]{Onuki_1994}%
  \BibitemOpen
  \bibfield  {author} {\bibinfo {author} {\bibfnamefont {A.}~\bibnamefont
  {Onuki}},\ }\bibfield  {title} {\bibinfo {title} {Domain growth and rheology
  in phase-separating binary mixtures with viscosity difference},\ }\href
  {https://doi.org/10.1209/0295-5075/28/3/004} {\bibfield  {journal} {\bibinfo
  {journal} {Europhys. Lett.}\ }\textbf {\bibinfo {volume} {28}},\ \bibinfo
  {pages} {175} (\bibinfo {year} {1994})}\BibitemShut {NoStop}%
\bibitem [{\citenamefont {Huo}\ \emph {et~al.}(2004)\citenamefont {Huo},
  \citenamefont {Zhang},\ and\ \citenamefont {Yang}}]{Huo2004}%
  \BibitemOpen
  \bibfield  {author} {\bibinfo {author} {\bibfnamefont {Y.}~\bibnamefont
  {Huo}}, \bibinfo {author} {\bibfnamefont {H.}~\bibnamefont {Zhang}},\ and\
  \bibinfo {author} {\bibfnamefont {Y.}~\bibnamefont {Yang}},\ }\bibfield
  {title} {\bibinfo {title} {Effects of reversible chemical reaction on
  morphology and domain growth of phase separating binary mixtures with
  viscosity difference},\ }\href
  {https://doi.org/https://doi.org/10.1002/mats.200300021} {\bibfield
  {journal} {\bibinfo  {journal} {Macromol. Theory Simul.}\ }\textbf {\bibinfo
  {volume} {13}},\ \bibinfo {pages} {280} (\bibinfo {year} {2004})}\BibitemShut
  {NoStop}%
\bibitem [{\citenamefont {Bauermann}\ \emph {et~al.}(2022)\citenamefont
  {Bauermann}, \citenamefont {Laha}, \citenamefont {McCall}, \citenamefont
  {J\"{u}licher},\ and\ \citenamefont {Weber}}]{bauermann2022chemical}%
  \BibitemOpen
  \bibfield  {author} {\bibinfo {author} {\bibfnamefont {J.}~\bibnamefont
  {Bauermann}}, \bibinfo {author} {\bibfnamefont {S.}~\bibnamefont {Laha}},
  \bibinfo {author} {\bibfnamefont {P.~M.}\ \bibnamefont {McCall}}, \bibinfo
  {author} {\bibfnamefont {F.}~\bibnamefont {J\"{u}licher}},\ and\ \bibinfo
  {author} {\bibfnamefont {C.~A.}\ \bibnamefont {Weber}},\ }\bibfield  {title}
  {\bibinfo {title} {Chemical kinetics and mass action in coexisting phases},\
  }\href {https://doi.org/https://doi.org/10.1021/jacs.2c06265} {\bibfield
  {journal} {\bibinfo  {journal} {J. Am. Chem. Soc.}\ }\textbf {\bibinfo
  {volume} {144}},\ \bibinfo {pages} {19294} (\bibinfo {year}
  {2022})}\BibitemShut {NoStop}%
\bibitem [{\citenamefont {Tanaka}(2022)}]{tanaka2022viscoelastic}%
  \BibitemOpen
  \bibfield  {author} {\bibinfo {author} {\bibfnamefont {H.}~\bibnamefont
  {Tanaka}},\ }\bibfield  {title} {\bibinfo {title} {Viscoelastic phase
  separation in biological cells},\ }\href
  {https://doi.org/10.1038/s42005-022-00947-7} {\bibfield  {journal} {\bibinfo
  {journal} {Commun. Phys.}\ }\textbf {\bibinfo {volume} {5}},\ \bibinfo
  {pages} {167} (\bibinfo {year} {2022})}\BibitemShut {NoStop}%
\bibitem [{\citenamefont {Tanaka}(2000)}]{Tanaka2000}%
  \BibitemOpen
  \bibfield  {author} {\bibinfo {author} {\bibfnamefont {H.}~\bibnamefont
  {Tanaka}},\ }\bibfield  {title} {\bibinfo {title} {Viscoelastic phase
  separation},\ }\href {https://doi.org/10.1088/0953-8984/12/15/201} {\bibfield
   {journal} {\bibinfo  {journal} {J Phys.: Condens. Matter}\ }\textbf
  {\bibinfo {volume} {12}},\ \bibinfo {pages} {R207} (\bibinfo {year}
  {2000})}\BibitemShut {NoStop}%
\end{thebibliography}%

\end{document}